\documentstyle[12pt]{article}

\def\Journal#1#2#3#4{{#1} {\bf #2}, #3 (#4)}

\def\NIM{\em Nucl. Instrum. Methods}

\def\PRL{\em Phys. Rev. Lett.}

\def\ea{{\it et al.}}

\def\beq{\begin{equation}}
\def\eeq{\end{equation}}
\def\beqa{\begin{eqnarray}}
\def\eeqa{\end{eqnarray}}

\def\sinw{\hbox{$\sin^{2}\theta^{\rm eff}_{W}$}}
\def\alr{A_{LR}}
\def\alrz{A_{LR}^{0}}
\def\Qlr{Q_{LR}}
\def\epem{e^{+}e^{-}}
\def\mpmm{\mu^{+}\mu^{-}}
\def\tptm{\tau^{+}\tau^{-}}
\def\Ae{A_{e}}
\def\Al{A_{l}}
\def\Am{A_{\mu}}
\def\At{A_{\tau}}
\def\Zz{Z^{0}}
\def\cv{v_{e}}
\def\ca{a_{e}}
\def\sigl{\sigma_{L}}
\def\sigr{\sigma_{R}}
\def\Pe{{\cal P}_{e}}
\def\Pg{{\cal P}_{\gamma}}

\renewcommand{\baselinestretch}{1.1}

\begin{document}


\thispagestyle{empty}


\begin{flushright}
{\footnotesize\renewcommand{\baselinestretch}{.75}
SLAC--PUB--7307\\
September 1996\\}
\end{flushright}

\begin{center}
{\bf\large Determination of Electroweak Parameters at the SLC$^*$}
\end{center}

\vspace{1.0cm}

\begin{center}
Eric Torrence\\
Massachusetts Institute of Technology\\
Cambridge, Massachusetts 02139\\
\bigskip

Representing the SLD Collaboration$^{\diamond}$\\
Stanford Linear Accelerator Center\\
Stanford University, Stanford, California 94039\\ 
\end{center}

\vfill

\begin{center} 
{\it Presented at the 28th International Conference on High Energy Physics}\\
{\it Warsaw, Poland}\\
{\it July 25-31, 1996}\\
\end{center}

\vfill

\noindent $^{*}${\footnotesize
 Work supported by Department of Energy contracts:
  DE-FG02-91ER40676 (BU),
  DE-FG03-91ER40618 (UCSB),
  DE-FG03-92ER40689 (UCSC),
  DE-FG03-93ER40788 (CSU),
  DE-FG02-91ER40672 (Colorado),
  DE-FG02-91ER40677 (Illinois),
  DE-AC03-76SF00098 (LBL),
  DE-FG02-92ER40715 (Massachusetts),
  DE-AC02-76ER03069 (MIT),
  DE-FG06-85ER40224 (Oregon),
  DE-AC03-76SF00515 (SLAC),
  DE-FG05-91ER40627 (Tennssee),
  DE-FG02-95ER40896 (Wisconsin),
  DE-FG02-92ER40704 (Yale);
 National Science Foundation grants:
  PHY-91-13428 (UCSC),
  PHY-89-21320 (Columbia),
  PHY-92-04239 (Cincinnati),
  PHY-88-17930 (Rutgers),
  PHY-88-19316 (Vanderbilt),
  PHY-92-03212 (Washington);
  the UK Science and Engineering Research Council
  (Brunel and RAL);
  the Istituto Nazionale di Fisica Nucleare of Italy
  (Bologna, Ferrara, Frascati, Pisa, Padova, Perugia);
  and the Japan-US Cooperative Research Project on High Energy Physics
  (Nagoya, Tohoku).
}\\
\noindent $^{\diamond}${\footnotesize
The SLD Collaboration authors and their institutions are listed 
following the references.
}
\eject




\renewcommand{\baselinestretch}{2}
\normalsize


\centerline{\bf Abstract}

We present an improved measurement of the left-right cross section
asymmetry ($\alr$) for $\Zz$ boson production by $\epem$ collisions.
The measurement was performed at a center-of-mass energy of 91.28 GeV
with the SLD detector at the SLAC Linear Collider (SLC) during the 
1994-95 running period.
The luminosity-weighted average polarization of the SLC electron beam
during this run was measured to be $(77.23 \pm 0.52)\% $.
Using a sample of 93,644 hadronic $\Zz$ decays, we measure the pole 
asymmetry $\alrz$ to be
$0.1512 \pm 0.0042({\rm stat.}) \pm 0.0011({\rm syst.})$
which is equivalent to an effective weak mixing angle of 
$\sinw = 0.23100 \pm 0.00054({\rm stat.}) \pm 0.00014({\rm syst.})$.
We also present a preliminary direct measurement of the 
$\Zz$-lepton coupling asymmetries $\Ae$,$\Am$, and $\At$ extracted
from the differential cross section observed in leptonic $\Zz$ decays.
We combine these results with our previous $\alr$ measurement to 
obtain a combined determination of the weak mixing angle
$\sinw = 0.23061 \pm 0.00047$. 
\vfill\eject

\section{Introduction}
\label{sec:alr-intro}

The left-right cross section asymmetry is defined as
$\alrz \equiv (\sigl - \sigr)/(\sigl+\sigr)$ where $\sigl$
and $\sigr$ are the $\epem$ production cross sections for $\Zz$
bosons at the $Z$ pole energy with left-handed and right-handed 
electrons, respectively.
The Standard Model predicts that this quantity depends upon the 
effective vector ($\cv$) and axial-vector ($\ca$) couplings of 
the $Z$ boson to the electron current,
\beq
\alrz = \frac{2 \cv \ca}{\cv^{2}+\ca^{2}} \equiv 
\frac{2 ( 1 - 4 \sinw ) }{ 1 + ( 1 - 4 \sinw )^{2}},
\label{eq:alr-alr0}
\eeq
where the effective weak mixing angle is defined as 
$\sinw \equiv (1-\cv/\ca)/4$.
Note that $\alrz$ is a sensitive function of $\sinw$ and depends upon 
virtual electroweak radiative corrections including those involving the
top quark and Higgs boson, as well as corrections arising from new 
phenomena.
The recent measurement of the top quark mass~\cite{topc,topd}
has greatly enhanced the
power of this measurement as a test of the prevailing theory.

The measurement is performed by counting the number of hadronic $Z$
decays observed for each of the two longitudinal polarization states 
of the incident electron beam.
A continual measurement of the electron beam polarization $\Pe$ allows
us to form the asymmetry 
\beq
\alr(E_{cm}) = \frac{1}{<\Pe>} \cdot \frac{N_{L}-N_{R}}{N_{L}+N_{R}}
\label{eq:alr-alr}
\eeq
where $<\Pe>$ is the luminosity-weighted electron beam polarization.
The experimental asymmetry $\alr$ must be corrected for small effects
arising from initial state radiation, pure photon exchange, and
$Z$-photon interference to extract $\alrz$.

\section{Electron Polarization at the SLC}
\label{sec:alr-slc}
At the SLC, longitudinally polarized electrons are produced by 
photoemission from a strained-lattice GaAs photocathode illuminated
by a Ti-Sapphire laser operating at 849 nm.\cite{source}
A Pockels cell driven by a pseudo-random sequence at 120 Hz determines
the helicity of the incident laser pulse and hence the helicity
of the produced electron bunch.
Optimization of the cathode design has improved the maximum beam 
polarization to nearly 80\%.

The spin transport of the electrons through the SLC remains unchanged 
from the 1993 run.
The polarization axis of the electron bunch is rotated into the
vertical plane before the damping ring, remains oriented vertically
during the acceleration phase, and is brought back into the horizontal
plane by means of a pair of large amplitude betatron oscillations
(``spin bumps'') in the SLC North arc.\cite{spin-bumps}
These spin bumps are empirically set to optimize the longitudinal
electron polarization at the SLD Interaction Point (IP).
The luminosity-weighted $\epem$ center-of-mass energy ($E_{cm}$)
is measured with precision energy spectrometers~\cite{wisrd}
to be $91.280 \pm 0.025$ GeV.
A detailed description of SLC operations with polarized electrons
can be found elsewhere.\cite{slc}

\section{The Compton Polarimeter}
The longitudinal electron beam polarization ($\Pe$) is measured
by a Compton scattering polarimeter located 33 meters
downstream of the IP.
A circularly polarized 2.33 eV photon beam produced by a frequency
doubled Nd:YAG laser is scattered off the exiting 45.6 GeV
electron bunch just before the beam enters the first set of dipole
magnets of the SLC South arc heading towards the electron beam dump.
These magnets act as a spectrometer sweeping the scattered
electrons out of the main SLC beam line and into a multichannel 
threshold Cherenkov detector where the momentum spectrum of the
electrons is measured in the interval from 17 to 30 GeV/c.

The counting rate in each detector channel is measured for parallel 
and anti-parallel combinations of the photon and electron beam 
helicities.
The asymmetry formed from these rates is equal to the product
$\Pe \Pg A(E)$ where $\Pg$ is the circular polarization of the laser
beam at the electron-photon crossing point and $A(E)$ is the 
theoretical asymmetry function at the accepted energy $E$ of the 
scattered electrons.

Polarimeter data are acquired continuously during the operation of
the SLC.
A statistical error of $\sim 1\%$ is reached in approximately three minutes,
although two thirds of the polarimeter data acquired are used for 
calibration purposes.
We obtain $\Pe$ from the observed asymmetry using the measured value 
of $\Pg$ and the theoretical asymmetry function (including $\sim 1\%$ 
corrections for detector resolution effects).
The systematic uncertainties associated with the polarization measurement,
summarized in Table~\ref{tab:polsys}, are currently dominated by our 
ability to measure the linearity of the entire polarimeter system.
For the 1994-95 run the total relative systematic uncertainty is 
estimated to be $\delta\Pe/\Pe = 0.67 \% $.

In our previous measurement based on data acquired in 1993,\cite{alr-93}
it was noted that the polarization measured by the 
Compton polarimeter does not exactly correspond to the polarization of 
the electrons producing $Z$ bosons at the SLD.
While the Compton polarimeter measures the polarization of the entire 
electron bunch, chromatic aberrations in the SLC final focus optics 
reduce the luminosity generated from the off-energy beam tails.
Because of the energy-dependent spin precession experienced by the 
electrons in the SLC North arc, these off-energy beam tails 
have a systematically lower net longitudinal polarization than the
beam core.

During the 1994-95 run, a number of measures were taken to control this
effect, both in the operation of the SLC and in monitoring procedures,
which have significantly reduced both the relative size of this effect
and the associated uncertainty from $(1.7\pm1.1)\%$ to $(0.20\pm0.14)\% $.
At this level, the spin precession of the electron bunch in the
final focus quadrupole triplet must also be taken into account, and
the net correction is included as an additional systematic 
uncertainty on the beam polarization measurement listed in 
Table~\ref{tab:polsys}.
In addition, depolarization due to the collision process itself was
directly measured and found to be negligible.

\begin{table}\begin{center}
\caption{Polarimeter Systematic Uncertainties}
\label{tab:polsys}
\vspace{0.4cm}
\begin{tabular}{lc}
\hline\hline
{\it Systematic Uncertainty} & $\delta\Pe/\Pe$(\%)\\
\hline
Laser Polarization & 0.20 \\
Detector Linearity & 0.50 \\
Detector Calibration & 0.29 \\
Electronic Noise & 0.20 \\
Compton - IP Difference & 0.17 \\
\hline
Total Uncertainty & 0.67 \\
\hline \hline
\end{tabular}
\end{center}
\end{table}

\section{Event Selection}

The $\epem$ collisions are measured by the SLD detector which has been
described elsewhere.\cite{slddet}
The triggering of the SLD relies upon a combination of calorimeter and
tracking information, and the event selection is based upon energy 
clusters reconstructed in the Liquid Argon Calorimeter 
(LAC),\cite{lac} and charged tracks reconstructed in the Central
Drift Chamber (CDC).\cite{cdc}
Cuts on minimum calorimeter energy and maximum calorimeter energy
imbalance are used to remove two-photon and beam related 
backgrounds, while a cut on minimum track multiplicity is used to 
remove $\epem$ final states.
These processes have a different left-right production asymmetry
than the hadronic $Z$ decays that we are interested in, and we 
apply a small correction to account for any residual background 
contamination in our sample.
The background in our event sample is estimated to be 
$(0.08 \pm 0.08)\% $ for $\epem$ final states and $(0.03 \pm 0.03)\% $
for the remainder.
Although they are not backgrounds, the other two leptonic final states 
are also selected with very poor efficiency: $\sim 7\%$ for $\tptm$,
$\sim 0\%$ for $\mpmm$.
We are left with a very pure sample of hadronic $Z$ decays with an 
estimated total efficiency of $(89\pm1)\%$.

\section{Measurement of $\bf \alr$}

A total of 93,644 events satisfy the selection criteria.
We find that 52,179 $(N_{L})$ and 41,465 $(N_{R})$ are produced from
the left-handed and right-handed electron helicity state respectively,
leading to a measured asymmetry $A_{m} = 0.11441 \pm 0.00325$.
After dividing by the luminosity weighted beam polarization,
measured to be $<\Pe>=(77.23 \pm 0.52)\%$, and applying
a correction of $\delta\alr/\alr = (0.240 \pm 0.055)\%$ to account for the 
residual background and other small beam asymmetries, we find the 
left right asymmetry at $E_{cm}$ = 91.28 GeV to be 
\[ \alr = 
0.1485 \pm 0.0042({\rm stat.}) \pm 0.0010({\rm syst.}).
\] 
Correcting this result for electroweak interference and initial state 
radiation, we find the pole asymmetry $\alrz$ to be
\[ \alrz = 
0.1512 \pm 0.0042({\rm stat.}) \pm 0.0011({\rm syst.}) ,
\]
which can be combined with our previous results~\cite{alr-92,alr-93}
to yield a cumulative value of
\beqa
\alrz & = & 0.1543 \pm 0.0039   \nonumber \\
\sinw & = & 0.23060 \pm 0.00050 \nonumber.
\eeqa

\section{Lepton Asymmetries}
While not nearly as precise as the hadronic $\alr$ measurement just
presented, additional information about the electroweak couplings
can be extracted by considering the leptonic decays of the $\Zz$ boson.
The polarized differential cross section for the process 
$\Zz \rightarrow l^{+}l^{-}$ can be written as
\beqa
\frac{d\sigma}{d\Omega} \propto
(1-\Pe \Ae) (1+\cos^{2}\theta) \nonumber
+2 (\Ae - \Pe) \Al \cos\theta ,
\label{eq:diffxsect}
\eeqa
where $\cos \theta$ is the production angle between the incoming
electron and outgoing lepton and $\Al$ is identical to the coupling
asymmetry defined in Equation~\ref{eq:alr-alr0} for a lepton of type 
$l$.
The $\epem$ final states have a more complicated form owing to the
additional t-channel photon exchange amplitude, and the analysis of 
these events will not be presented here.
For the remaining $\mpmm$ and $\tptm$ events, the quantities $\Ae$,
$\Am$, and $\At$ can be extracted by performing an unbinned likelihood
fit to the observed data within a fiducial tracking region of
$|\cos\theta| < 0.7$.

Events are selected with a number of tracking and calorimetric cuts
to reject $\epem$, two photon, and hadronic final states.
A total of 3,788 $\mpmm$ events are selected with an estimated
efficiency of 95\% in the fiducial tracking region, and an estimated
background contamination of 0.4\% primarily from the tau final states.
A sample of 3,748 $\tptm$ events are selected with a slightly lower
efficiency of 89\% in the fiducial region and a slightly higher
estimated background of 4\% mostly from the $\mpmm$ final states,
along with $\sim 1\%$ $\epem$ contamination.
These events have been selected from the combined 1993-95 SLD data set.
A number of systematic uncertainties related to the background
determination, angular acceptance estimate, and electroweak 
interference correction have been considered and all found to be negligible
compared to the statistical error of each measurement.
The results, shown in Table~\ref{tab:numbers}, are competitive with
the forward-backward asymmetry measurements performed by a single
LEP experiment with a factor of $\sim 20$ more data.
This analysis is not complete, and the results from the lepton
final states presented in Table~\ref{tab:numbers} are preliminary.

Also listed in Table~\ref{tab:numbers} are two other measurements
of $\Ae$ performed by the SLD collaboration, which are included for
completeness.
The first, $\Ae(\rm bhabha)$, is an old measurement of the $\epem$
final states based on 1992-93 data only.\cite{pitts}
The second, $Q_{LR}$, is a measurement of the left-right asymmetry in
the inclusive hadronic charge flow.\cite{baranko}
Since the uncertainty in each of the measurements listed in 
Table~\ref{tab:numbers} is dominated by statistics, the correlated
systematic uncertainty due to the electron polarization
measurement is negligible and can be safely ignored.
Statistical correlations between the various analyses have
been estimated to be no larger than $\sim 7 \%$ and can also
be safely ignored.

\begin{table}\begin{center}
\caption{Lepton Asymmetry Measurements}
\label{tab:numbers}
\vspace{0.4cm}
\begin{tabular}{cc}
\hline\hline
{\it Method} & {\it Result} \\
\hline
Leptonic     & $\Ae = 0.148 \pm 0.016$ \\
Final States & $\Am = 0.102 \pm 0.033$ \\
({\em Preliminary}) & $\At = 0.190 \pm 0.034$ \\
\hline
$\Ae({\rm bhabha})$ & $\Ae = 0.202 \pm 0.038$  \\
$\Qlr$ & $\Ae = 0.162 \pm 0.043$ \\
$\alr$ & $\Ae = 0.1543 \pm 0.0039$ \\
\hline
Combined & $\Al = 0.1542 \pm 0.0037$ \\
\hline \hline
\end{tabular}
\end{center}
\end{table}

\section{Conclusions}

We note that the measurement of $\alr$ presented here represents the
single most precise determination of $\sinw$ by a single experiment.
If we assume the universality of the lepton current coupling 
parameters, we can include the preliminary results from the
lepton final states to produce a slightly improved
combined SLD result of
\beqa
\Ae   & = & 0.1542  \pm 0.0037  \nonumber \\
\sinw & = & 0.23061 \pm 0.00047 \nonumber.
\eeqa
It should be noted that the uncertainty on this result is
dominated by statistics.
The SLD has been approved to run into the year 1998, and 
we expect to be able to reduce the uncertainty on $\sinw$ by 
another factor of two.

\section*{Acknowledgments}

We thank the personnel of the SLAC accelerator department and the 
technical staffs of our collaborating institutions for their 
outstanding efforts on our behalf.

\vfill\eject

\bigskip
\begin{center}
%
%
  \def\iADEL{$^{(1)}$}
  \def\iBOL{$^{(2)}$}
  \def\iBU{$^{(3)}$}
  \def\iBRUN{$^{(4)}$}
  \def\iUCSB{$^{(5)}$}
  \def\iUCSC{$^{(6)}$}
  \def\iCIN{$^{(7)}$}
  \def\iCSU{$^{(8)}$}
  \def\iCOLO{$^{(9)}$}
  \def\iCOL{$^{(10)}$}
  \def\iFER{$^{(11)}$}
  \def\iFRA{$^{(12)}$}
  \def\iILL{$^{(13)}$}
  \def\iLBL{$^{(14)}$}
  \def\iMIT{$^{(15)}$}
  \def\iMASS{$^{(16)}$}
  \def\iMISS{$^{(17)}$}
  \def\iMOSC{$^{(18)}$}
  \def\iNAG{$^{(19)}$}
  \def\iOREG{$^{(20)}$}
  \def\iPAD{$^{(21)}$}
  \def\iPERU{$^{(22)}$}
  \def\iPISA{$^{(23)}$}
  \def\iRUT{$^{(24)}$}
  \def\iRAL{$^{(25)}$}
  \def\iSOGANG{$^{(26)}$}
  \def\iSLAC{$^{(27)}$}
  \def\iTENN{$^{(28)}$}
  \def\iTOH{$^{(29)}$}
  \def\iVAND{$^{(30)}$}
  \def\iWASH{$^{(31)}$}
  \def\iWISC{$^{(32)}$}
  \def\iYALE{$^{(33)}$}
  \def\dead{$^{\dag}$}
  \def\andgen{$^{(a)}$}
  \def\andper{$^{(b)}$}
  \baselineskip=.75\baselineskip   
  \advance\leftskip by -0.5cm        
  \advance\rightskip by -0.5cm

{\bf\large The SLD Collaboration} \\
\bigskip
\mbox{K. Abe                 \unskip,\iNAG}
\mbox{K. Abe                 \unskip,\iTOH}
\mbox{I. Abt                 \unskip,\iILL}
\mbox{T. Akagi               \unskip,\iSLAC}
\mbox{N.J. Allen             \unskip,\iBRUN}
\mbox{W.W. Ash               \unskip,\iSLAC$^\dagger$}
\mbox{D. Aston               \unskip,\iSLAC}
\mbox{K.G. Baird             \unskip,\iRUT}
\mbox{C. Baltay              \unskip,\iYALE}
\mbox{H.R. Band              \unskip,\iWISC
% ...
}
\it
  \vskip \baselineskip
  \iADEL
     Adelphi University,
     Garden City, New York 11530 \break
  \iBOL
     INFN Sezione di Bologna,
     I-40126 Bologna, Italy \break
  \iBU
     Boston University,
     Boston, Massachusetts 02215 \break
  \iBRUN
     Brunel University,
     Uxbridge, Middlesex UB8 3PH, United Kingdom \break
  \iUCSB
     University of California at Santa Barbara,
     Santa Barbara, California 93106 \break
  \iUCSC
     University of California at Santa Cruz,
     Santa Cruz, California 95064 \break
  \iCIN
     University of Cincinnati,
     Cincinnati, Ohio 45221 \break
  \iCSU
     Colorado State University,
     Fort Collins, Colorado 80523 \break
  \iCOLO
     University of Colorado,
     Boulder, Colorado 80309 \break
  \iCOL
     Columbia University,
     New York, New York 10027 \break
  \iFER
     INFN Sezione di Ferrara and Universit\`a di Ferrara,
     I-44100 Ferrara, Italy \break
  \iFRA
     INFN  Lab. Nazionali di Frascati,
     I-00044 Frascati, Italy \break
  \iILL
     University of Illinois,
     Urbana, Illinois 61801 \break
  \iLBL
     Lawrence Berkeley Laboratory, University of California,
     Berkeley, California 94720 \break
  \iMIT
     Massachusetts Institute of Technology,
     Cambridge, Massachusetts 02139 \break
  \iMASS
     University of Massachusetts,
     Amherst, Massachusetts 01003 \break
  \iMISS
     University of Mississippi,
     University, Mississippi  38677 \break
  \iMOSC
     Moscow State University,
     Institute of Nuclear Physics,
     119899 Moscow,
     Russia    \break
  \iNAG
     Nagoya University,
     Chikusa-ku, Nagoya 464 Japan  \break
  \iOREG
     University of Oregon,
     Eugene, Oregon 97403 \break
  \iPAD
     INFN Sezione di Padova and Universit\`a di Padova,
     I-35100 Padova, Italy \break
  \iPERU
     INFN Sezione di Perugia and Universit\`a di Perugia,
     I-06100 Perugia, Italy \break
  \iPISA
     INFN Sezione di Pisa and Universit\`a di Pisa,
     I-56100 Pisa, Italy \break
  \iRUT
     Rutgers University,
     Piscataway, New Jersey 08855 \break
  \iRAL
     Rutherford Appleton Laboratory,
     Chilton, Didcot, Oxon OX11 0QX United Kingdom \break
  \iSOGANG
     Sogang University,
     Seoul, Korea \break
  \iSLAC
     Stanford Linear Accelerator Center, Stanford University,
     Stanford, California 94309 \break
  \iTENN
     University of Tennessee,
     Knoxville, Tennessee 37996 \break
  \iTOH
     Tohoku University,
     Sendai 980 Japan \break
  \iVAND
     Vanderbilt University,
     Nashville, Tennessee 37235 \break
  \iWASH
     University of Washington,
     Seattle, Washington 98195 \break
  \iWISC
     University of Wisconsin,
     Madison, Wisconsin 53706 \break
  \iYALE
     Yale University,
     New Haven, Connecticut 06511 \break
  \dead
     Deceased \break
  \andgen
     Also at the Universit\`a di Genova \break
  \andper
     Also at the Universit\`a di Perugia \break
\rm


\end{center}

\end{document}